\documentclass[useAMS,usenatbib,aas_macros,onecolumn]{mn2e}

\usepackage{epsfig}
\usepackage{lscape,graphicx,colortbl}
\usepackage{amssymb}
\usepackage{natbib}
\usepackage{times}
\usepackage{aas_macros}

\newcommand{\vFWHM}{\ifmmode V_{\mbox{\tiny FWHM}} \else $V_{\mbox{\tiny FWHM}}$ \fi}

\newcommand{\ld}{\ifmmode {\rm lt-days} \else lt-days \fi}
\newcommand{\kms}{\ifmmode {\rm km\,s}^{-1} \else km\,s$^{-1}$ \fi}
\newcommand{\cc}{\hbox{cm$^{-3}$}}

\newcommand{\ergs}{\ifmmode {\rm erg\,s}^{-1} \else erg\,s$^{-1}$ \fi}
\newcommand{\ergcms}{\ifmmode {\rm erg\,cm}^{-2}\,{\rm s}^{-1} \else erg\,cm$^{-2}$\,s$^{-1}$\fi}
\newcommand{\ergcmsA}{\ifmmode{\rm erg}\, {\rm cm}^{-2}\,{\rm s}^{-1}\,{\rm\AA}^{-1} \else erg\, cm$^{-2}$\, s$^{-1}$\, \AA$^{-1}$\fi}
\newcommand{\ergcmsHz}{\ifmmode{\rm erg\,cm}^{-2}\,{\rm s}^{-1}\,{\rm Hz}^{-1} \else erg\,cm$^{-2}$\,s$^{-1}$\,Hz$^{-1}$\fi}
\newcommand{\phcms}{\ifmmode {\rm ph\,cm}^{-2}\,{\rm s}^{-1} \else ,ph\,cm$^{-2}$\,s$^{-1}$\fi}
\newcommand{\phcmsA}{\ifmmode {\rm ph\,cm}^{-2}\,{\rm s}^{-1}\,{\rm\AA}^{-1} \else ph\,cm$^{-2}$\,s$^{-1}$\,\AA$^{-1}$\fi}

\newcommand{\Lsun}{\ifmmode L_{\odot} \else $L_{\odot}$\fi}
\newcommand{\auvo}{\ifmmode \alpha_{\nu,{\rm UVO}} \else $\alpha_{\nu,{\rm UVO}}$\fi}
\newcommand{\Luv}{\ifmmode L_{1450} \else $L_{1450}$\fi}
\newcommand{\Lop}{\ifmmode L_{5100} \else $L_{5100}$\fi}
\newcommand{\Lthree}{\ifmmode L_{3000} \else $L_{3000}$\fi}
\newcommand{\lledd}{\ifmmode L/L_{\rm Edd} \else $L/L_{\rm Edd}$\fi}
\newcommand{\Ledd}{\ifmmode L/L_{\rm Edd} \else $L/L_{\rm Edd}$\fi}
\newcommand{\lamLlam}{\ifmmode \lambda L_{\lambda} \else $\lambda L_{\lambda}$\fi}
\newcommand{\Lbol}{\ifmmode L_{\rm bol} \else $L_{\rm bol}$\fi}
\newcommand{\lLbol}{\ifmmode \log\left(\Lbol/\ergs\right) \else $\log\left(\Lbol/\ergs\right)$\fi}

\newcommand{\Fthree}{\ifmmode F_{3000} \else $F_{3000}$\fi}

\newcommand{\Halpha}{\ifmmode {\rm H}\alpha \else H$\alpha$\fi}
\newcommand{\ha}{\ifmmode {\rm H}\alpha \else H$\alpha$\fi}
\newcommand{\Lya}{\ifmmode {\rm Ly}\alpha \else Ly$\alpha$\fi}
\newcommand{\Hbeta}{\ifmmode {\rm H}\beta \else H$\beta$\fi}
\newcommand{\hb}{\ifmmode {\rm H}\beta \else H$\beta$\fi}
\newcommand{\MgII}{\ifmmode {\rm Mg}\,\textsc{ii}\,\lambda2798 \else Mg\,{\sc ii}\,$\lambda2798$\fi}
\newcommand{\mgii}{\ifmmode {\rm Mg}{\textsc{ii}} \else Mg\,{\sc ii}\fi}
\newcommand{\CIV}{\ifmmode {\rm C}\,\textsc{iv}\,\lambda1549 \else C\,{\sc iv}\,$\lambda1549$\fi}
\newcommand{\civ}{\ifmmode {\rm C}\,\textsc{iv} \else C\,{\sc iv}\fi}

\newcommand{\oi}{\ifmmode \left[{\rm O}\,\textsc{i}\right] \else [O\,{\sc i}]\fi}
\newcommand{\OI}{\ifmmode \left[{\rm O}\,\textsc{i}\right]\,\lambda6300 \else [O\,{\sc i}]$\,\lambda6300$ \fi}

\newcommand{\oii}{\ifmmode \left[{\rm O}\,\textsc{ii}\right] \else [O\,{\sc ii}]\fi}
\newcommand{\OII}{\ifmmode \left[{\rm O}\,\textsc{ii}\right]\,\lambda3727 \else [O\,{\sc ii}]\,$\lambda3727$ \fi}
\newcommand{\oiii}{\ifmmode \left[{\rm O}\,\textsc{iii}\right] \else [O\,{\sc iii}]\fi}
\newcommand{\OIII}{\ifmmode \left[{\rm O}\,\textsc{iii}\right]\,\lambda5007 \else [O\,{\sc iii}]\,$\lambda5007$\fi}

\newcommand{\Lnu}{\ifmmode L_{\rm \nu} \else $L_{\rm \nu}$\fi}
\newcommand{\Llambda}{\ifmmode L_{\rm \lambda} \else $L_{\rm \lambda}$\fi}
\newcommand{\LA}{\ifmmode L_{\rm 5100} \else $L_{\rm 5100}$\fi}
\newcommand{\Lg}{\ifmmode L_{\rm g(r)} \else $L_{\rm g(r)}$\fi}
\newcommand{\Ln}{\ifmmode L_{\rm n(r)} \else $L_{\rm n(r)}$\fi}
\newcommand{\Lo}{\ifmmode L_{\rm 0} \else $L_{\rm 0}$\fi}
\newcommand{\Ro}{\ifmmode R_{\rm 0} \else $R_{\rm 0}$\fi}
\newcommand{\Rhalf}{\ifmmode R_{\rm 1/2} \else $R_{\rm 1/2}$\fi}
\newcommand{\Rg}{\ifmmode R_{\rm g} \else $R_{\rm g}$\fi}
\newcommand{\Rms}{\ifmmode R_{\rm ms} \else $R_{\rm ms}$\fi}
\newcommand{\Rin}{\ifmmode R_{\rm in} \else $R_{\rm in}$\fi}
\newcommand{\Rout}{\ifmmode R_{\rm out} \else $R_{\rm out}$\fi}
\newcommand{\RISCO}{\ifmmode R_{\rm in} \else $R_{\rm in}$\fi}
\newcommand{\ro}{\ifmmode r_{\rm 0} \else $r_{\rm 0}$\fi}
\newcommand{\rhalf}{\ifmmode r_{\rm 1/2} \else $r_{\rm 1/2}$\fi}
\newcommand{\rg}{\ifmmode r_{\rm g} \else $r_{\rm g}$\fi}
\newcommand{\rms}{\ifmmode r_{\rm ms} \else $r_{\rm ms}$\fi}
\newcommand{\rin}{\ifmmode r_{\rm in} \else $r_{\rm in}$\fi}
\newcommand{\rout}{\ifmmode r_{\rm out} \else $r_{\rm out}$\fi}
\newcommand{\rISCO}{\ifmmode r_{\rm in} \else $r_{\rm in}$\fi}
\newcommand{\Fnu}{\ifmmode F_{\nu} \else $F_{\nu}$\fi}
\newcommand{\Flambda}{\ifmmode F_{\lambda} \else $F_{\lambda}$\fi}
\newcommand{\Mdot}{\ifmmode \dot{M} \else $\dot{M}$\fi}
\newcommand{\mdot}{\ifmmode \dot{m} \else $\dot{m}$\fi}
\newcommand{\Mrdot}{\ifmmode \dot{M}\left(r \right) \else $\dot{M}\left(r \right)$\fi}
\newcommand{\MRdot}{\ifmmode \dot{M}\left(R \right) \else $\dot{M}\left(R \right)$\fi}
\newcommand{\Mrindot}{\ifmmode \dot{M}\left(r_{ISCO} \right) \else $\dot{M}\left(r_{ISCO} \right)$\fi}
\newcommand{\Mroutdot}{\ifmmode \dot{M}\left(r_{out} \right) \else $\dot{M}\left(r_{out} \right)$\fi}
\newcommand{\MBHdot}{\ifmmode \dot{M}_{BH} \else $\dot{M}_{BH}$\fi}
\newcommand{\MBHdotexpct}{\ifmmode \dot{M}_{BHexpected} \else $\dot{M}_{BHexpected}$\fi}
\newcommand{\Medddot}{\ifmmode \dot{M}_{edd} \else $\dot{M}_{edd}$\fi}
\newcommand{\Moutdot}{\ifmmode \dot{M}_{out} \else $\dot{M}_{out}$\fi}
\newcommand{\Mindot}{\ifmmode \dot{M}_{in} \else $\dot{M}_{in}$\fi}
\newcommand{\Mwinddot}{\ifmmode \dot{M}_{wind} \else $\dot{M}_{wind}$\fi}
\newcommand{\MBH}{\ifmmode M_{\rm BH} \else $M_{\rm BH}$\fi}
\newcommand{\mbh}{\ifmmode M_{\rm BH} \else $M_{\rm BH}$\fi}
\newcommand{\Mexp}{\ifmmode M_{\rm 8} \else $M_{\rm 8}$\fi}

\newcommand{\Msun}{\ifmmode M_{\odot} \else $M_{\odot}$\fi}
\newcommand{\msun}{\ifmmode M_{\odot} \else $M_{\odot}$\fi}
\newcommand{\avisc}{\ifmmode \alpha_{visc} \else $\alpha_{visc}$\fi}

\title[Growth rate and slim accretion discs in active galactic nuclei]
{Bolometric luminosity black-hole growth time and slim accretion discs in active galactic nuclei}

\author[Hagai Netzer, Benny Trakhtenbrot]
{Hagai Netzer $^1$\thanks{E-mail: netzer@wise.tau.ac.il} and Benny Trakhtenbrot $^{2,3}$\\
$^1$School of Physics and Astronomy, The Sackler Faculty of Exact Sciences, Tel-Aviv University, Tel-Aviv 69978, Israel\\
$^2$Department of Particle Physics and Astrophysics, The Weizmann Institute of Science, Rehovot, 76100 Israel 
(Benoziyo postdoctoral fellow)\\
$^3$Institute for Astronomy, Department of Physics, ETH Zurich, Wolfgang-Pauli-Strasse 27, CH-8093 Zurich, Switzerland
(Zwicky postdoctoral fellow)
} 

\date{Submitted 2013}

\begin{document}

\maketitle



\begin{abstract}
\noindent
We investigate the accretion rate, bolometric luminosity, black hole (BH) growth time and BH spin in a large AGN sample
under the assumption that all such objects are powered via thin or slim accretion discs (ADs). We use direct estimates 
of the mass accretion rate, \Mdot, to show that many currently used values of \Lbol\ and \Ledd\ are 
either under estimated or over estimated because they are based on bolometric correction factors that are adjusted to the properties 
of moderately accreting  
active galactic nuclei (AGN) and do not take into account the correct combination of BH mass, spin and accretion rate.
The consistent application of AD physics  
to our sample of Sloan Digital Sky Survey (SDSS) AGN leads to the following findings:
1. Even the most conservative assumption about the radiative efficiency of fast accreting BHs shows that 
many of these sources must contain slim ADs. We illustrate this by estimating the fraction of such objects at various redshifts. 
2. Many previously estimated BH growth times are inconsistent with the AD theory. In particular, the growth times
of the fastest accreting BHs were over estimated in the past by large factors with important 
consequences to AGN evolution. 
3. Currently used bolometric correction factors for low accretion rate very massive SDSS BHs, are inconsistent with the AD theory. 
Applying the AD set of assumptions to such objects, combined with standard photoionization calculations of broad emission lines,
leads to the conclusion that many such objects must contain fast spinning BHs.

\end{abstract}


\section{Introduction}
\label{sec:introduction}

Super massive black holes (BHs) in active galactic nuclei (AGN) are thought to be powered by mass accretion trough various types of accretion flows. 
Cold gas that originates in the host galaxy, or a nearby companion, contains enough angular momentum to result in a non-spherical flow. This, as well as direct observational evidence, lead to the suggestion of a central small accretion disc (AD) that facilitates the loss of angular momentum, and eventual feeding of the BH, through various types of viscosity.
Much of the theoretical study of such flows, and the comparison of their spectral energy distribution (SED) with spectroscopic observations of AGN, have been focused on geometrically thin, optically thick accretion discs, those systems with $H/R \ll 1$, where $H$ is the thickness of the disc at radius $R$. 
Examples are the seminal work of \citet[][hereafter SS73]{SS73} and numerous publications that are listed in the 
comprehensive review by \cite{Koratkar1999} and the more recent paper by \cite{DavisLaor2011}.

Recently, there has been a renewed interest in standard and non-standard AGN ADs. This  includes the suggestion of very cold accretion discs around very massive BHs \citep{LaorDavis2011}, winds and outflows from thin discs \citep{Slone2012} and study of truncated, two components discs whose main source of UV radiation is diluted gas inside the classical SS73 disc \citep{Done2012}.
``Slim accretion discs'' have also been considered in the context of BH feeding. These  systems are characterized by fast accretion rates and high \Ledd, which results in radiation pressure dominating over gas pressure and a thick geometry.  
Such systems can differ substantially from thin accretion discs \citep[e.g.,][and references therein]{Abramowicz1988,Mineshige2000,Wang2003,Ohsuga2011}. 
In particular, the escape time of the photons emitted in such flows,
close to the BH, can become longer than the accretion time, leading to advection and reduction in the mass-to-radiation conversion efficiency.
This is usually referred to as the ``saturated luminosity'' of slim ADs. In addition, a
hot corona can develop over a large part of the surface of of a slim disc and the overall SED can differ substantially 
from the thin disc SED.
The result is a significant modification in the well defined relationships between accretion rate, BH mass and BH spin predicted by the 
thin AD theory.
A recent work by \cite{Wang2013} argued that extreme slim discs, with $\Ledd \gg 1$, can be used as a new type of ``cosmological candles'' since their bolometric luminosity, \Lbol, is directly proportional to the BH mass with only a weak (logarithmic) dependence on the accretion rate. 
 Additional properties of such systems, as well as the basic terminology used in the study of such systems, are explained in the recent study of Pu et al. (2013)
Despite their very intriguing physical properties, and the potential observational consequences, the fraction of slim ADs among AGN 
remains unknown and poorly studied

This work investigates the accretion rate, bolometric luminosity, growth time and spin of active BHs in large AGN samples.
We start from the basic assumption that all active BHs are powered by ADs whose mass accretion rate, \Mdot, is given by the standard thin-disc approximation \citep[e.g.,][]{DavisLaor2011}. 
We then study the implications to active BH in large AGN samples which we divide according to their BH mass (\mbh) and redshift.
\S2\ describes the calculations of \Lbol\ and \Ledd\ under different assumptions about AGN discs. 
In \S3 we apply these calculations to a large sample of Sloan Digital Sky Survey (SDSS) AGN and
show that even the most conservative assumptions about the radiative efficiency lead to the conclusion that the fraction of slim ADs  must be large, at all redshifts. 
We also investigate the growth times of active BHs and show why previous estimates of this time might have been wrong.
Finally we make a connection between low accretion rate in high mass BHs and the BH spin.
\S4 summarizes the results and the major conclusions of this paper.
Throughout this work we assume a cosmological model with $\Omega_{\Lambda}=0.7$, $\Omega_{m}=0.3$, and
$H_{0}=70\,\kms\,\, {\rm Mpc}^{-1}$.

\section{Calculations}

The goal of this work is to explore a broad range of physical conditions that determine  \Lbol\ and  \Ledd\ in AGN with different BH mass and accretion rate. 
This is done under the assumption that the entire continuum spectral energy distribution (SED), except for the hard X-ray regime, is emitted by a central thin or slim AD. We assume that all the energy is provided by mass accretion and re-radiated by the AD.
There are certainly some exceptions, especially among nearby low luminosity AGN where the X-ray luminosity is a large part
of \Lbol. This is true for only a small fraction of the population considered here and will be neglected
for the rest of the paper.

Most present day AD models are based on the original blackbody (BB) thin-disc model of SS73 with various modifications, such as a more accurate general relativity (GR) treatment and improved radiative transfer in the disc atmosphere \citep[e.g.,][]{Hubeny2001,DavisLaor2011}. 
Such modifications are important when comparing a detailed theoretical disc SED to real observations. 
As is evident from composite AGN spectra such as the ones presented in \cite[e.g.,][]{Vandenberk2001,Richards2006}, a large number of AGN SEDs cannot be explained by simple (BB) AD models. The additional modifications, especially those related to the radiation transfer in the disc atmosphere, can improve the agreement considerably. 
In particular, there are good reasons to assume (see below) that the global energetics of of such systems, as well as \Mdot, can be reliably estimated by such models once the BH spin ($a$) is taken into account.

Our work makes use of thin AD calculations. The numerical code we use is the one described in \cite{Slone2012}. It is based on the SS73 set of assumptions with a fixed viscosity parameter $\alpha$ (taken here to be 0.1) and a spin-dependent innermost stable circular orbit (ISCO) which determines the mass-to-radiation conversion efficiency $\eta$. 
The model incorporates a full relativistic treatment for all spin parameters ($a$) larger than zero and  takes into account Comptonization of the emitted radiation in the disc atmosphere. For $a<0$, it neglects GR effects but not the atmospheric Comptonization. This is a fair approximation
since for such spins, the ISCO is far enough from the BH (between 6 and 9 gravitational radii).
The novel ingredient of the model is the ability to include the effect of mass outflow from the surface of the disc. 
The current application of the model, however, assumes no disc wind.

For thin ADs, and long enough wavelengths, the mass accretion rate, bolometric luminosity, monochromatic luminosity ($L_{\nu}$), or monochromatic observed flux ($F_{\nu}$),are related thorough various known expressions derived by \cite{Collin2002}, 
\cite{DavisLaor2011} and several others\footnote{Note that \cite{DavisLaor2011} contains an error that was corrected in \cite{LaorDavis2011}}.
We express them here in a somewhat modified way, based on \cite{Netzer2013_book}:
\begin{equation}
4 \pi D_L^2 F_{\nu} =f(\theta) [M_8 \dot{M_{\odot}} ]^{2/3} \left[ 
  \frac{\lambda}{5100{\rm \AA}} 
  \right]^{-1/3} 
   \,\, \ergcmsHz  \,\, ,
\label{eq:5100}
\end{equation}
where $M_8$ is the BH mass in units of $10^8\,\msun$, $\dot{M_{\odot}}$ the accretion rate in units of \Msun/yr, and $D_L$ the luminosity distance. The inclination dependent term, $f(\theta)$, includes the normalization of the emitted flux
 and the angular dependence of the emitted radiation. It can be expressed as: 
\begin{equation}
 f(\theta) = \frac{f_0 F_{\nu}}{F_{\nu}(\textrm{face-on})} = f_0 \frac {\cos i (1+ b(\nu) \cos i)}{1+b(\nu)}  \,\, ,
\label{eq:limb}
\end{equation}
where $i$ is the inclination to the line of sight and  $b(\nu)$ is a limb darkening function which, in many situations, can be expressed in a frequency independent 
manner, e.g. $b(\nu)\simeq 2$.
The constant $f_0$ depends on $b(\nu)$. For $b(\nu)=2$,  $f_0 \simeq 1.2 \times 10^{30} ~ {\rm erg/sec/Hz}$ 
and for $b(\nu)=0$,  $f_0 \simeq 9.5 \times 10^{29} ~ {\rm erg/sec/Hz}$.
All SEDs shown in this work are for face-on discs ($\cos i=1$) with a frequency independent limb-darkening factor of $b=2$.

Throughout this work we use the normalized (or Eddington)  mass accretion rate defined by 
\begin{equation}
 \mdot= \frac{\Mdot}{\Mdot_{Edd}} \, ,
\label{eq:mdot}
\end{equation}
where $\Lbol=\eta \Mdot c^2$ and $L_{\rm Edd}=\eta \Mdot_{\rm Edd} c^2$. We assume $L_{\rm Edd}=1.5x10^{46}\,M_8\,\ergs$ which applies to solar composition gas. Using this definition, \mdot=\Ledd.  Throughout the work we use \mdot\ rather than \Ledd.

\begin{figure}
\centering
\includegraphics[width=9cm]{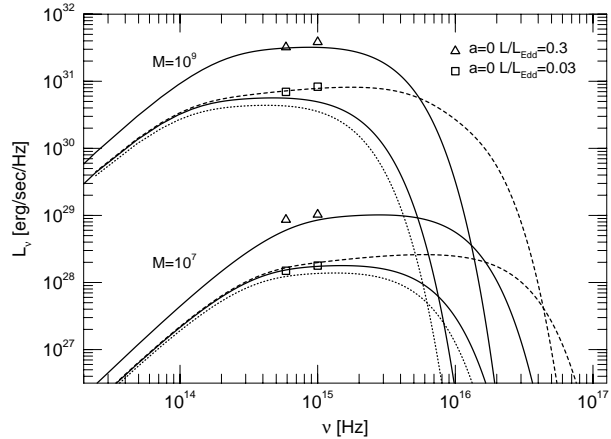}
\caption{
Thin disc SEDs for \mbh=$10^7$ and $10^9$\msun\ and various accretion rates. 
The solid lines are theoretical spectra for $a=0$ for each of the BHs, one for \Mdot\ that gives $\mdot=0.3$ and one that gives $\mdot=0.03$. The two calculated values of accretion rates (Eqn.~\ref{eq:5100}) are marked with squares ($\mdot=0.03$) and triangles ($\mdot=0.3$). Dashed and dotted lines show theoretical SEDs with $a=0.998$ and $a=-1$, respectively, for two values of \Mdot\ that give $\mdot=0.03$ for $a=0$ (and hence larger \mdot\ for $a=0.998$ and smaller \mdot\ for $a=-1$).
}
\label{fig:5100_method}
\end{figure}

\cite{DavisLaor2011} showed that estimates of \Mdot\ based on expressions like Eq.~\ref{eq:5100} are in good agreement with the more detailed AD calculations provided the comparison of observed and calculated $L_{\nu}$ is made at a long enough wavelength (which they took to be at 4861\AA).
We demonstrate this in Figure~\ref{fig:5100_method} using our own calculations for two BH masses, $10^7$ and $10^9$ \msun, and several values of the spin parameter: $a=0$ (solid lines), $a=0.998$ (dashed lines) and $a=-1$ (dotted lines).
For each \mbh\ we show two calculated spectra with different normalized accretion rates, \mdot=0.03 (squares) and \mdot=0.3 (triangles). 
We chose two representing wavelengths that are used in many studies of AGN: 5100\AA\ and 3000\AA\ (for more information  see  the figure caption).
As expected, the deviations from the simple analytical prediction of Eqn.~\ref{eq:5100} increase with increasing \mbh\ and with decreasing wavelength. 
The diagram shows that using the flux at 5100\AA, the estimated \Mdot\ is in reasonable agreement with the calculations even for very massive BHs with $\mbh \sim 10^9\,\msun$. The deviations can be larger at 3000\AA, in particular for the case of the lowest spin, $a=-1$. 

The basic assumption in \cite{DavisLaor2011} is that all AGN are powered by thin ADs whose mass accretion rate can be determined from the value of \Mdot\ derived from continuum observations at 
4861\AA. 
They then integrated over the observed and extrapolated SED and obtained the resulting \Lbol\ for every object in a large sample
of PG quasars. The combination of \Mdot\ and \Lbol\ was used by them to estimate $\eta$. The results were used to claim a strong correlation between $\eta$ and \mbh\ in this sample.
Here we take an orthogonal approach that is also based on thin ADs but treats $\eta$ as a real unknown and uses the estimated \Mdot\
to probe the {\it entire} allowed  range of \Lbol\ in a large AGN sample. We then compare these estimates to
various empirical methods used in the literature to obtain \Lbol\ and hence \mdot.

Empirical methods to estimate \Lbol\ from optical-UV observations are discussed in numerous papers, most
recently in, e.g., \citet[][hereafter TN12]{Trak2012}, \cite{Runnoe2012} and \cite{KellyShen2013}. 
TN12 adopted the luminosity dependent bolometric correction factors of \cite{Marconi2004}. 
These bolometric correction factors are marked here by $b_{\lambda}$, e.g. $b_{5100}$. 
They convert $\lambda L_{\lambda}$ at a chosen wavelength to \Lbol.
The 5100\AA\ bolometric correction factor used by TN12 can be approximated by a simple expression that takes into account the fact that it decreases logarithmically with \Lop. 
Thus, to a good approximation,
\begin{equation}
b_{5100}=53-\log \Lop \, .
\label{eq:b_5100}
\end{equation}
Other factors can be approximated in a similar way, e.g. $b_{1400}=0.5 b_{5100}$, etc.
The correction factor used by \cite{KellyShen2013} is adjusted to rest-frame 2500\AA\ with a luminosity independent value of $b_{2500}=5$. 
Given a typical SED for luminous AGN, this is very similar to the other expressions given here. 

The above estimates should be compared with theoretical bolometric correction factors for thin ADs. 
Such a comparison is shown in Table~\ref{tab:bol_corr}, where we list theoretical, AD-based values of $b_{5100}$, alongside the empirical expression of Eq.~\ref{eq:b_5100}, for the various models shown in Fig.~\ref{fig:5100_method}. 
All numbers are calculated for face-on discs.
The table shows that in  some cases, the theoretical factors are similar to the factors given in Eq.~\ref{eq:b_5100} (after allowing for disc inclination). In others, the deviations are very significant.
The last column in Table~\ref{tab:bol_corr} lists the fraction of the  Lyman continuum luminosity, $L({\rm E}>13.6~{\rm eV})/ \Lbol$, emitted by such discs. 
As shown in \S3, this can be linked to BH spin in large mass, low accretion rate BHs.

\begin{table}
\centering
\caption{Face-on bolometric correction factors for the disc models shown in Fig.~\ref{fig:5100_method} }
\begin{tabular}{lcccccc}
\mbh/\msun\ & $a$   & \mdot\ &log \Lop\ (erg/sec) & $b_{5100}$(model) &$b_{5100}$(Eqn.~\ref{eq:b_5100})& $L({\rm E}>13.6~{\rm eV})/\Lbol$\\
\hline
$10^7$      & 0     &  0.03    & 42.95           &4.87       & 10         & 0.56   \\
$10^7$      & 0.998 &  0.03    & 42.99           & 15.98     & 10          & 0.85 \\
$10^7$      & -1    &  0.03    & 42.86           & 4.56      & 10.1        & 0.53 \\
$10^7$      & 0     &  0.3     &  43.54          & 12.35     & 9.4         & 0.76   \\
$10^9$      & 0     &  0.03    & 45.52          &1.35       &  7.5        & 0.068 \\
$10^9$      & 0.998 &  0.03    & 45.62          &3.98       & 7.5         & 0.49 \\
$10^9$      & -1    &  0.03    & 45.40          &1.26       &  7.6        &  0.050   \\
$10^9$      & 0     &  0.3     & 46.26          &2.40       & 6.7         & 0.28 \\
\end{tabular}
\label{tab:bol_corr}
\end{table}

\section{Results and discussion}
\subsection{The ``correct'' \Lbol\ and \mdot}

Having introduced the calculations and their limitations, we now apply them to several large AGN samples in order to examine three issues:
the range of expected \Lbol\ and \mdot\ under the assumption of thin ADs, the consequence to BH growth times and spin, and the fraction of AGN where the thin disc approximation no longer applies. 

In what follows we are \emph{not} interested in the absolute number of AGN with different properties. 
Because of this, our estimates do not take into account well known selection effects, especially near the flux limit of the SDSS sample \citep[see, e.g.,][]{KellyShen2013,Wu2013_SDSS_epsilon}, that are of fundamental importance for calculating various cosmological properties like the luminosity and mass functions of AGN. 
We are only interested in the often-neglected ways by which the AD physics may affect the derived properties of the AGN population. 
When discussing such properties, we assume that BH spin, and therefore $\eta$, are completely unknown.

The sample we analyze is the SDSS sample presented in TN12. It is based
on the SDSS/DR7 data set \citep{Abazajian2009} and includes a large number of sources in the redshift range 0.1--2.   
All BH mass measurements were obtained with the single epoch (or ``virial'') method - a well established method based on several 
successful reverberation mapping (RM) experiments. 
TN12 presented such mass estimates based on measured \Lop\ and FWHM(\hb) ($z<0.75$), or measured \Lthree\ and FWHM(\MgII) ($z>0.5$). 
Here we focus on three redshift groups, $z=0.1-0.25$ (FWHM[\hb]-based mass estimates), $z= 0.6-0.75$ (FWHM[\hb]-based mass estimates), and $z=1.6-1.75$ (FWHM[\MgII]-based mass estimates), and two mass groups, $\mbh\simeq 10^8$ and $10^9$ \msun.
The mass groups are defined to include sources where \mbh\ is within $\pm0.15$ dex of the chosen mass.
The specific expressions that are used to calculated the mass are Eqs.~3 and 12 in TN12. 
We exclude sources with $\Lop<2 \times 10^{43}\,\ergs$ to avoid complications due to host galaxy contamination. This amounts to the omission of a few percent of the sources in the lower redshift groups. 

\begin{figure}
\center
\includegraphics[width=0.7\textwidth]{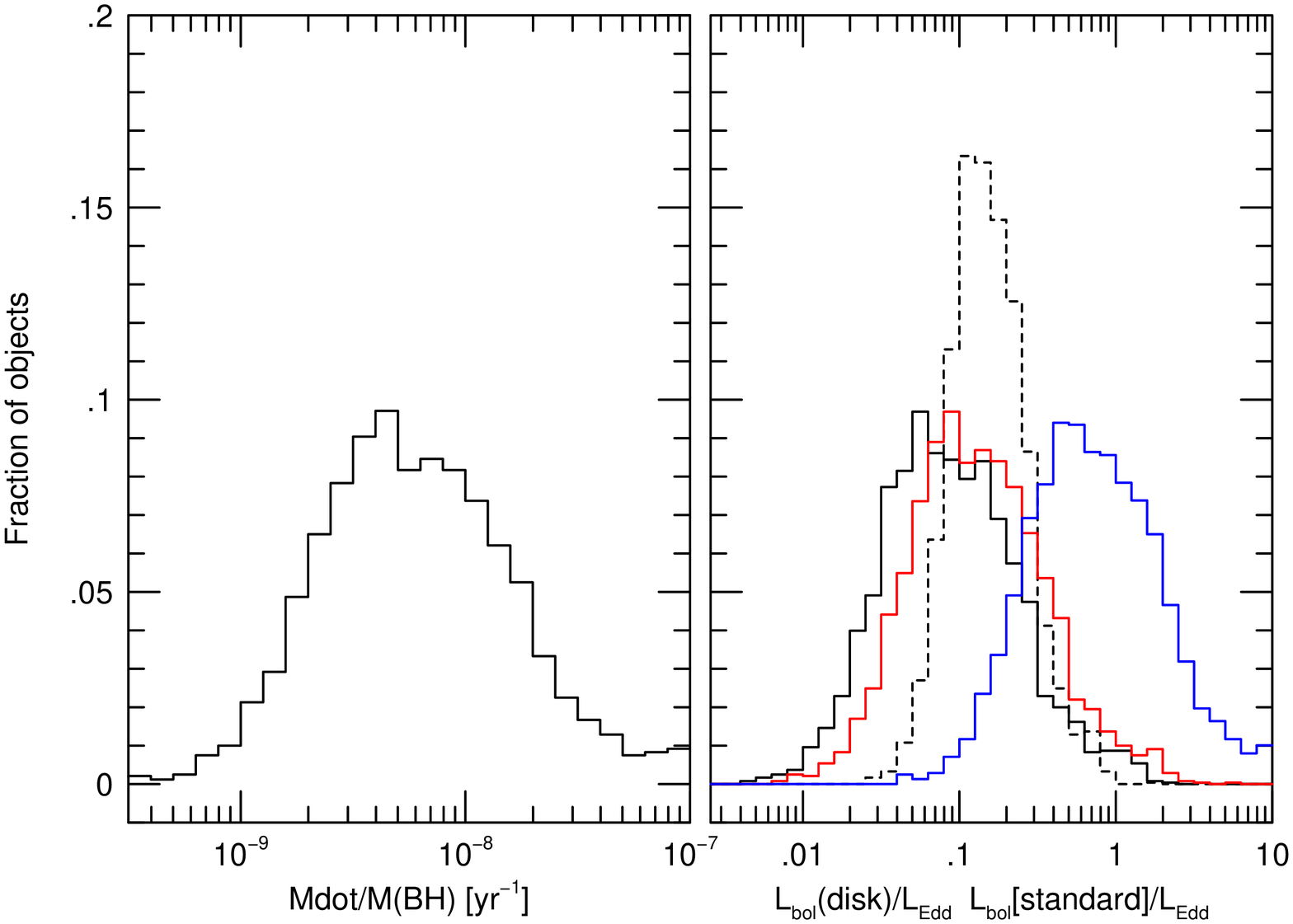}
\caption{
{\it Left:} The distribution of \Mdot/\mbh\ in the group of sources with $\mbh\simeq 10^8\,\msun$ at $z= 0.6-0.75$.
{\it Right:} Four derived \mdot\ (\Ledd) distributions for the sources in the left panel using various assumptions about BH 
spin and bolometric correction factors. Black, $a=-1$, 
red, $a=0$ and blue, $a=0.998$.
The standard assumptions of TN12 using \Lop\ and the bolometric correction factor
of \citet{Marconi2004} is shown in a black dashed line.
\label{fig:Ledd_comparison}
}
\end{figure}

The measured \Lop, \Lthree\ and \mbh\ in TN12 were used to derive \Mdot\ based on Eq.~\ref{eq:5100}. 
We then obtained \Mdot/\mbh, \Lbol(disc) (which we refer to in other parts of the paper as simply \Lbol)
 and \mdot\ for three BH spins: $a=0$ (stationary BH, $\eta=0.057$), $a=-1$ (retrograde disc, $\eta=0.038$) and $a=0.998$ (maximally rotating BH, $\eta=0.32$).
Fig.~\ref{fig:Ledd_comparison} shows the distributions of these properties for the group of $z=0.6-0.75$ and $\mbh\simeq 10^8\, \msun$ sources, 
and compares them with the distribution obtained from calculating \Lbol\ 
by the method of TN12, who used the empirical bolometric correction factors of \cite{Marconi2004} (which we 
coin \Lbol[standard]). 
Equivalently, we define $\mdot \left[{\rm standard}\right]\equiv\Lbol\left[{\rm standard}\right]/1.5 \times 10^{46}\,M_8$. 
As seen from the diagram, a single well-defined distribution of \Mdot/\mbh\ results in a large range of possible \mdot\ due to the uncertain value of $\eta$ and hence \Lbol.

\begin{figure}
\center
\includegraphics[width=0.7\textwidth]{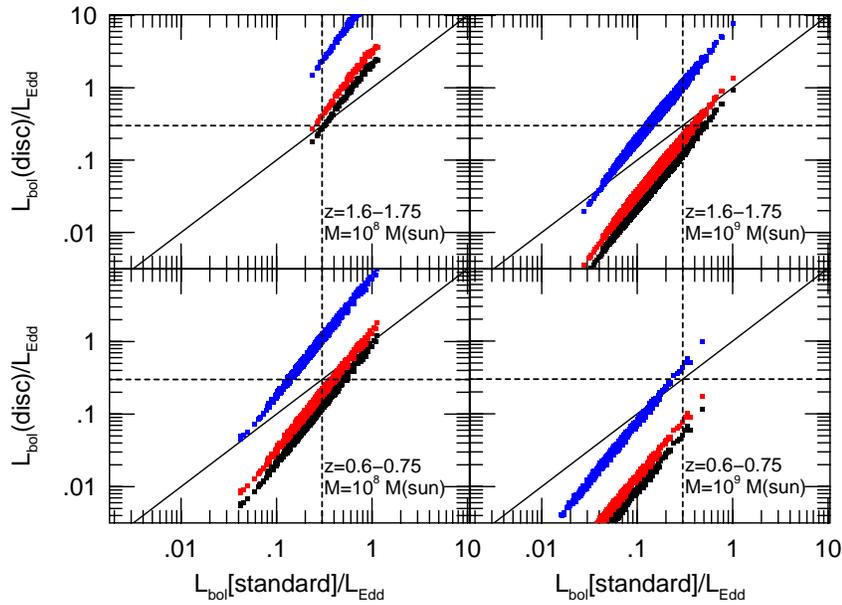}
\caption{Various spin-dependent values of \mdot\ vs. \mdot[standard] computed with the TN12 method.
We show four redshift and \mbh\ groups, as marked. The 1:1 ratios are shown by solid lines in all panels. 
Colors represent three spin parameters:
black for $a=-1$, red for $a=0$ and blue for $a=0.998$. 
Note that for some cases (lower left and upper right panels) the various methods are in reasonable agreement
with each other while for others (lower right and upper left panels) the disagreement can be very large.
\mdot=0.3 is marked by dashed lines in all panels to emphasize
the potential transition from thin to slim ADs (see text). Note that all values of $\mdot>0.3$ are calculated assuming
thin ADs and are therefore likely to over-estimate the actual value because of the saturated luminosities of slim ADs (see text).
\label{fig:Ledd_benny_vs_disc}
}
\end{figure}

Figure~\ref{fig:Ledd_benny_vs_disc} presents the range and uncertainty on \Lbol\  in a somewhat different way. The diagram compares 
\mdot[standard] to the value of \mdot\ obtained by directly measuring \Mdot\ and assuming thin ADs with various assumed spin parameters (vertical axis).
Large deviations from \mdot[standard], in some cases by up to an order of magnitude,  are evident in all groups.
The reason is that for a given BH mass, the standard assumption assumes bolometric correction factors that are only weakly dependent on \Lop\ or \Lthree\ (e.g. Eq.~\ref{eq:b_5100}). 
The thin disc model predicts a very different dependence of the bolometric correction factors which, for a given spin 
is proportional to  $\left(\mdot/\mbh\right)^{1/3}$.
The result is that in every mass group, there is a range of \Lbol\ (and therefore \mdot) where the two estimates agree, mostly because the bolometric correction factors in use are based on real observations of the most common AGN (usually those with $\mdot\left({\rm standard}\right)\sim 0.1$). 
However, each group also contains numerous sources that deviate substantially from this simplified calibration. 
As explained in \S1, the plotted values along the vertical axis are likely to over-estimate for \mdot$>0.3$
because of the saturated luminosities of slim ADs.
As we show below, there are important consequences to BH growth times in cases where the AD-based value of
\mdot\ exceeds \mdot[standard]. There are different consequences to BH spin in the other
extreme, where \mdot[standard] exceeds the value obtained from the thin AD theory.

\subsection{Thin or slim accretion discs}

According to \cite{Laor1989}, the thin disc approximation breaks down when \mdot\ exceeds about 0.3.
More recent studies, based on an improved solution of the vertical structure of the disc \citep[e.g.][and references therein]{Sadowski2011}), 
reach similar conclusions and show the dependence of the limiting \mdot\ on the BH spin. 
The limiting accretion rates in those calculations are in the range $0.1< \mdot <0.5$. Here we are not interested in the details 
of the slim disc model (described in \S~\ref{sec:introduction}) but rather in the fraction of such objects in the AGN population. 
Because of this we adopt a single value of $\mdot=0.3$ as marking the transition from thin to slim ADs.

Figure~\ref{fig:Ledd_benny_vs_disc} illustrates the large number of slim ADs at low and high redshifts.
This is shown by horizontal and vertical lines marking the boundary of $\mdot=0.3$ measured by the different methods.
Evidently, there is a large number of such objects in three of the four panels of the diagram.
More quantitative data are given in Table~\ref{tab:fraction_slim_discs}, where we list the 
fractions of objects with $\mdot>0.3$ for various  groups of BH mass and spin at two representing redshifts.
The fraction of slim AGN discs increases with redshift which is not surprising given that
the mean \mdot\ for all size BHs in the SDSS and similar AGN samples 
increases with redshift up to at least $z=2$ (TN12 and references therein).
 
So far we have considered the spin dependent value of \mdot\ which determine the transition from thin to slim discs.
However, we are more interested in the cases where BH accretion {\it must proceed}  via slim ADs. To obtain this, we
use the {\it most conservative assumption} on the BH spin, $a=-1$, which will result in the smallest possible \mdot. 
If this assumption results in \mdot$>0.3$, then the disk must be slim.
We find that even this conservative estimate  results in a large fractions of slim ADs at all redshifts and most
BH mass.  For example, Table~\ref{tab:fraction_slim_discs} shows that at $z= 1.6-1.75$ {\it at least} 93\% of SDSS-detected BHs with $\mbh\sim\ 10^8\,\mbh$ are suspected to be powered by slim ADs.

\begin{table}
\centering
\caption{The fraction of slim accretion discs in various BH mass and redshift groups
(the word ``standard'' means that we use \mdot[standard] as defined in the text). }
\begin{tabular}{lccc}
\mbh/\msun\ & $z$         &  $a$    & Fraction  \\
\hline
$10^8$      & 06--0.75    &  -1       & 0.13   \\
$10^8$      & 1.6--1.75   &  -1       & 0.93  \\
$10^8$      & 0.6--0.75   &   0       & 0.23  \\
$10^8$      & 1.6--1.75   &   0       & 0.99\\
$10^8$      & 0.6--0.75   &  0.998    & 0.91    \\
$10^8$      & 1.6--1.75   &  0.998    & 1.0\\
$10^9$      & 1.6--1.75   &  -1       & 0.01    \\
$10^9$      & 1.6--1.75   &  0        & 0.02 \\
$10^9$      & 1.6--1.75   &  0.998    & 0.56\\
\hline
$10^8$      & 1.6--1.75   & standard  & 0.95   \\
$10^9$      & 1.6--1.75   &  standard & 0.06\\
\end{tabular}
\label{tab:fraction_slim_discs}
\end{table}

As explained in \S1, SEDs of slim ADs are likely to deviate significantly from those
of thin ADs. In particular, the radiative efficiency $\eta$ is smaller due to photon trapping and has been estimated to behave like $\eta \propto$\mdot$^{-1}$.
This makes slim ADS dimmer compared with model predictions based on thin disks with the same \Mdot. There are other consequences regarding the X-ray radiation
(i.e., steeper than $\Gamma_{\rm X}=2$), the radiation pattern, and the bolometric correction factor $b_{5100}$ which, above a certain \mdot, no longer depends on BH spin and accretion rate. 
While all these issues are not very well understood theoretically, it is expected that the short wavelength SED of such objects will deviate, considerably, from the one predicted by the thin disk theory. 
Future studies of AGN will have to take these considerations into account when calculating the luminosity function of AGN, BH growth time, BH spin, the ionization of the gas around the active BH, and more. 
Here we only comment on two of these issues, namely BH growth times and spins.
In the rest of this paper we continue to use the thin disc definition of  
$\eta$ but note that beyond $\mdot\sim 0.3$, this represents an upper limit to the radiative efficiency. 

\subsection{The growth time of massive black holes}
The directly measured \Mdot\ provides the most meaningful way to estimate the typical growth time of active BHs at
various epochs. We demonstrate this by considering exponential growth, i.e. constant \Mdot/\mbh.
A common expression used to compute the
continuous growth time, (i.e. the growth time that does not including the activity duty cycle) is
$t_{\rm growth} \propto  \eta/(1-\eta)$/\mdot\  
with an additional term which depends logarithmically on the seed (beginning of growth) BH mass.
A big uncertainty in calculating this time is the need to use \mdot[standard] instead of \mdot\ which is not known because of
the unknown $\eta$. This introduces a large uncertainty due to the combination of 
the unknown bolometric correction factor and $\eta/(1-\eta)$. Our calculations show that the combined effect can 
easily amounts to a factor of 5 uncertainty in $t_{\rm growth}$.
The assumption of a thin AD, and the directly measured value of \Mdot/\mbh, results in a more accurate expression,
\begin{equation}
 t_{\rm growth} = \frac {1/(1-\eta) }{ \Mdot/\mbh} \ln \left[ \frac{\mbh }{ M_{\rm seed} } \right] \, {\rm yr} \, ,
\label{eq:t_grow}
\end{equation}
which removes a big part of the uncertainty, because the range in $1/(1-\eta)$ is a factor of 1.7 at most.
It is interesting to study the changes in $t_{\rm growth}$ resulting from the use of this more accurate expression compared to the results of the older calculations.

\begin{figure}
\center
\includegraphics[width=0.7\textwidth]{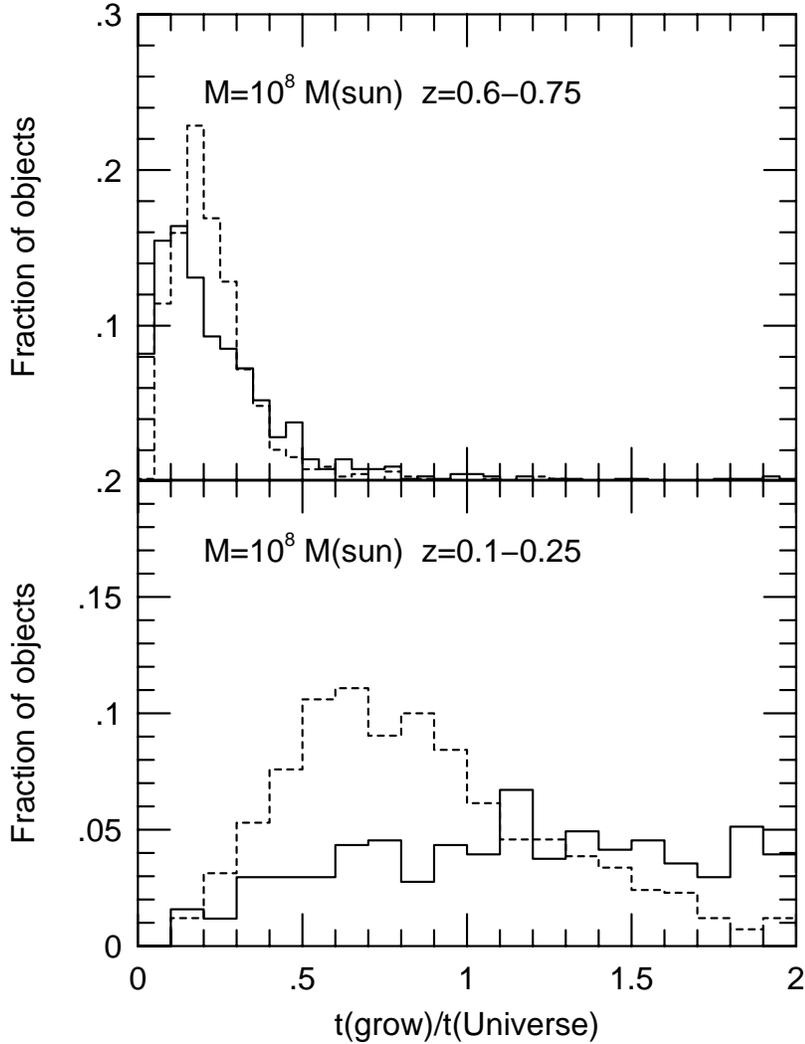}
\caption{
New (based on Eqn.~\ref{eq:t_grow}, solid line) versus old (``standard method'', 
dashed line) calculations of  continuous growth
times relative to the age of the universe for \mbh=$10^8$\,\msun.
{\it Bottom:} $0.1<z<0.25$. {\it Top:} $0.6<z<0.75$.
Note that the standard method under-estimates the
growth times of most BHs at low redshift and over-estimates them at higher redshift.
}
\label{fig:t_grow}
\end{figure}

Fig.~\ref{fig:t_grow} shows two computed distributions of $t_{\rm growth}/t({\rm Universe})$, where $t({\rm Universe})$ is the age of the universe at the redshift of the object, for two redshift intervals, $z= 0.1- 0.25$ and $z= 0.6-0.75$. 
In both cases \mbh=$10^8$\msun\ and $M_{\rm seed}=10^4$\msun\ and for the standard case, $\eta=0.1$..
The diagram illustrates the differences between the new calculations, based on the known \Mdot/\mbh, and those based on the older method with \mdot[standard]. The results depend strongly on the redshift since all BHs with a given mass, including those chosen in
this example, accrete faster at higher redshifts. In the lower redshift bin, on the bottom panel, \Mdot\ is relatively
small and the actual, more accurate growth time is longer
than what is obtained by using \mdot[standard] (i.e. growth times based on \mdot[standard] are shorter than they really are).
In the higher redshift group, the mean \Mdot\ is larger and the \mdot[standard]-based 
growth times over-estimate the real $t_{\rm growth}$\footnote{The typical \Mdot\ in the group of \mbh$\simeq 10^8$\Msun/yr
and $z=0.6-0.75$ is between 0.5 and 2 \msun/yr}. Thus, previous growth time  estimates for many BHs with large accretion rates
must have been over-estimated in the past, especially at redshift larger than about 0.6.

An interesting case concerns the most challenging example; the
largest BHs at the highest redshift where there is less time to reach the observed \mbh\ through continuous
exponential growth. Our calculations show that for the sample in question, with $z \le 2$
and \mbh$\sim 10^9$\msun, 
the two methods give similar $t_{\rm growth}$. The reason is that such BHs are characterized
by a relatively small \Mdot/\mbh, even at $z \sim 2$.

\subsection{\Mdot\ \mdot\ and BH spin}
The previous discussion emphasized those cases where \Lbol[standard] under-estimates
the real \Lbol\ for much of the population. However, there are cases where the opposite is true and mass accretion is so slow
that it raises a sever question about the validity of \Lbol[standard].

Consider the slowest accreting, most massive BHs. Our AD-based calculations show
that  for a significant number of such sources \Lbol[standard] exceeds 
$\eta \Mdot c^2$ {\it regardless} of what value of spin we assume. 
These objects are the ones with the lowest \Mdot/\mbh\ in each mass and redshift bin and their fraction is higher among larger BHs. An example is shown in Table 1 where all possible spins
in the \mbh=10$^9$\msun, \mdot=0.03 group result in AD-based bolometric correction factors that are smaller than the standard
$b_{5100}$.
 Either accretion through a thin AD is not a viable model for such objects or else
the standard method of estimating bolometric luminosities breaks down.
We suspect the latter is correct because standard bolometric correction
factors, like  the ones derived by \cite{Marconi2004}, are based on observations of a small number of sources with 
accretion rates typical of AGN discovered in large and shallow samples. In most such cases,
$\mdot\sim 0.1$, i.e. an order of magnitude or more larger than the accretion rates of the massive
BHs in question. 
As explained, the standard $b_{5100}$ or $b_{1400}$ do not depend on BH mass while in thin ADs with a given BH spin, the bolometric correction factor depends on (\mdot/\mbh)$^{1/3}$. At large \mbh\ and small \mdot, the difference between the two estimated values of \Lbol\ can be significant.
Evidently, the use of standard bolometric correction factors has its limitation and extreme cases of very massive, slowly accreting BHs represent one group of sources where such approximations fail.

Assume now that all very massive, slowly accreting BHs in our sample are indeed powered through thin ADs.
These sources are included in the SDSS sample because of their strong and broad emission lines. This means that the fraction of the total luminosity emitted in the hydrogen Lyman continuum, $L({\rm E}>13.6~{\rm eV})$/\Lbol, must exceed a certain value or else the emission lines would be too weak to detect.
There have been numerous studies of this fraction based on the comparison of the combined 
energy of all broad emission lines and the estimate of the covering fraction ($C_{\rm f}$) of the ionizing source by high density 
material in the BLR \citep[e.g.][]{Netzer1985b,Collin1986,Korista1997a}. Here we address these issues in relation to the slow accreting, high mass BHs.

Standard BLR modelling suggests that $C_{\rm f} \sim 0.1$ with a likely range of 0.1--0.3 
\citep[see][and references therein]{Netzer2013_book}. 
We can use this estimate to calculate the equivalent width (EW) of the strongest broad lines in a BLR which is ionized by the radiation of a central AD. We have carried out several such calculations using the photoionization code ION \citep[see e.g.][]{NetzerMarziani2010} and assuming various SEDs typical of thin ADs
for a range of BH mass and accretion rates, and for various ionization parameters. The EW of all strong emission lines were calculated and the possible detection by SDSS-like sample evaluated taken into account the redshift (which determines the observed wavelength range and hence the lines most likely to be detected). 
As an example, consider a BH with $\mbh=10^9\,\msun$ and $L\left({\rm E>}13.6~{\rm eV}\right)/\Lbol =0.1$ in the center of a BLR with $C_{\rm f}=0.3$, which we consider to be close to the upper limit of the covering factor distribution. 
We assume very weak X-ray (0.1--100 keV) emission, that does not contribute more than 3\% to the total radiation. This assumption is in agreement with the
known decrease in the relative luminosity of an X-ray power-law with \Lbol.
For an ionization parameter of $10^{-1.5}$ and gas density of $10^{10}$\cc, typical of the assumed conditions over most of the BLR, we obtain 
EW(\ha)$\sim 65$\AA, 
EW(\hb)$\sim 10$\AA, 
EW(\mgii)$\sim 3$\AA, 
EW(${\rm C} \textsc{iii}]\lambda 1908+{\rm Si} \textsc{iii}]\lambda 1895$) $\sim 4$\AA\ 
and 
EW(\Lya)$\sim 40$\AA, 
all in the rest-frame of the source. 
These numbers are insensitive to the exact value of the ionization parameter or the gas density. 

For a virialized BLR gas, FWHM(line)$ \propto [M/R_{\rm BLR}]^{1/2} \propto M^{1/2} L_{5100}^{-\alpha/2}$, where $\alpha \sim 0.5-0.7$ is the slope of the 
$R_{\rm BLR}-L$ relationship. 
Inserting known constants we get 
FWHM(\hb)$\simeq 1700\,\left(M_{\rm BH}/10^8\,M_{\odot}\right)^{1/2} \left(\Lbol/10^{46}\,\ergs\right)^{-1/4}$ \kms\ 
\citep[see][eqn. 7.22]{Netzer2013_book}.
The typical numbers for the objects in question, with large \mbh\ and low \mdot, is FWHM(\hb)$\sim 12,000$ \kms.
Such low EW, very broad emission line objects, will not be identified as type-I AGN by SDSS-like surveys unless at very low redshift, 
to detect \ha, or at high enough redshift to detect \Lya. Thus, many low accretion rate high mass BHs are not going to be discovered unless
$L\left(E>13.6~{\rm eV}\right)/L_{\rm bol} > 0.1$.

Table~\ref{tab:bol_corr} lists the fraction of ionizing radiation in thin AD models with various BH masses, accretion rates and spins. 
The table shows the strong dependence of $L({\rm E}>13.6~{\rm eV})/$\Lbol\ on BH spin typical of such models. 
The most extreme cases with the lowest \mdot\ are indeed below the cut-off required to produce lines that are strong enough (assuming $C_{\rm f}<0.3$) to be detected by SDSS-like samples. To further illustrate this point, we consider the
group of SDSS sources with $\mbh\simeq 10^9\,\msun$ and $z = 0.6-0.75.6$ and calculated the minimum value of \mdot\
which determines this ionization limit in 4 spin groups, $a=-1$, ,$a=0$, $a=0.7$ and $a=0.998$. The 
 case of $a=0.7$ was chosen because it corresponds to $\eta=0.104$, close to the value of 0.1
used in various studies about AGN and BH evolution.

\begin{figure}
\centering
\includegraphics[width=9cm]{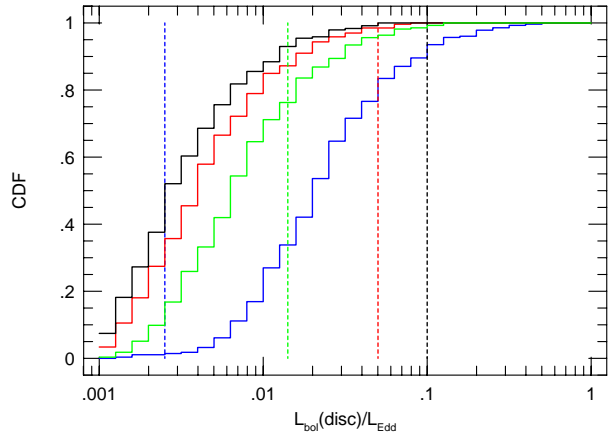}
\caption{
Cumulative density functions (CDFs) of \mdot\ for various spin groups of active BHs
with $\mbh\simeq 10^9\,\msun$ at $z=0.6-0.75$. The spins are marked by different colors:
Black: $a=-1$,. red: $a=0$, green $a=0.7$ and blue $a=0.998$.
Vertical lines with colors corresponding to color of their group mark the location of the critical \mdot. All
objects to the left of the line do not have a large enough Lyman continuum luminosity 
to produce detectable broad emission lines (see text for more explanation). 
}
\label{fig:lyman_fraction}
\end{figure}

We calculated \mdot\ and $L({\rm E}>13.6~{\rm eV})/\Lbol$ for all sources and 
constructed cumulative density functions (CDFs) of \mdot\ for each of the spin groups. The results are shown in Fig.~\ref{fig:lyman_fraction}.
The four vertical lines in the diagram, with colors corresponding to the spin group colors, mark the value of \mdot($a$) (hereafter ``limiting \mdot'') corresponding to $L({\rm E}>13.6~{\rm eV})/\Lbol=0.1$. 
All sources to the left of the line in the spin group in question {\it do not accrete fast enough} to produce the minimally
required fraction of Lyman continuum luminosity. The percentage of these sources are:
100\% for $a=-1$ (limiting $\mdot=0.1$), 
98\% for $a=0$ (limiting $\mdot=0.05$),  
76\% for $a=0.7$ (limiting $\mdot=0.014$) and 
4\% for $a=0.998$ (limiting $\mdot=0.0025$). 
The results of this simple statistical analysis suggest that in this mass and redshift group,
more than 98\% of the sources must have $a>0$ and more than 76\% must have $a > 0.7$. 
Thus, most of these BHs spin very fast, with spin parameters very close to the highest possible
value of $a=0.998$. 

The present analysis is not meant to identify specific BHs with large spin parameters. 
Such attempts have been made in other studies, mainly by detailed analysis of individual sources that show signs of relativistically broadened Fe-K$\alpha$ lines or objects with unusual radio properties.
These attempts have revealed a large range in BH spins, but did not result in clear relations between the BH spin and other AGN properties \cite[see, e.g.,][and references therein]{Reynolds2013}.
Instead, we point out the properties of a certain group of SDSS sources,
identified as type-I AGN, and argue, on statistical grounds, that many of them must contain fast rotating BHs.
We can think of two possible explanations for the unusual spin distribution of these sources.  
One is related to their large mass that can hint to the final stages of the growth process. It suggests 
that in most active, very massive BHs, the accretion rate goes down in time towards the end of the final accretion episode. 
According to this scenario,  we are witnessing the last accretion episode which lasted long enough to spin-up the BH close to the maximum possible speed. 
The second, perhaps more likely possibility is related to selection effects. The assumption is that the general population of large BH  includes both fast spinning and slow spinning BHs. 
However, the broad emission lines of the slowly spinning sources are not strong enough to allow their detection by the SDSS and similar sensitivity spectroscopic samples. 

Finally we mention the population of ``lineless AGN'', those sources with a strong AGN-like continuum
but extremely weak emission lines \citep[so-called WLQs; see, e.g.,][and references therein]{Fan1999, DiamondStanic2009, Shemmer2009}. 
Of the various explanations proposed to these intriguing sources, the one suggested by \cite{LaorDavis2011} emphasizes the possibility that the lines are 
weak and very broad because the BH mass is large and the accretion rate too small to produce significant ionizing continuum radiation. 
This is naturally related to the large mass, low accretion rate objects considered
here. However, our approach is different and based on the likelihood of detecting such
sources in big spectroscopic sample. In addition, \cite{LaorDavis2011} did not investigate the global spin and accretion rate distribution of such objects.

\section{Summary and conclusions}
The calculations presented here are based on the assumption that all AGN are powered by accretion through thin or slim ADs.
We follow the (slightly modified)  \cite{DavisLaor2011} recipe to estimate  \Mdot\ for many SDSS AGN with measured \mbh\
but unlike their work, we used
the AD theory to deduce the acceptable range of \Lbol\ and \Ledd\ (or \mdot) in these objects. 
We make use of the lowest allowed value of $\eta$ to obtain a conservative lower limit on \mdot\ which
enable us to draw the line between thin and slim ADs. This is then used to estimate
the fraction of slim ADs in the general AGN population. Finally, we looked at the two extremes of the \mdot\ 
distribution and investigated the consequences to BH growth times and BH spin. 
The main results can be summarized as followed:
\begin{enumerate}
 \item 
 The  distributions of \Lbol\ and \mdot\ in the AGN population is broader and more uncertain than assumed so far.
This is the result of the large uncertainties associated with currently-used bolometric correction factors 
that are based on objects that are more easily detected by large spectroscopic samples, and do not take into account
the large range of possible BH spins.
 The AD assumption leads to the conclusion that many AGN emit considerably 
more energy than currently assumed. 
\item
 Slim ADs, defined here as those systems with $\mdot>0.3$, are very common among AGN at all redshifts and most BH mass 
 groups. For example, the fraction of such AGN among SDSS sources at $z=1.6-1.75$ with \mbh$\sim 10^8$\msun, 
 is larger than 90\%. In all such cases, the actual \Lbol\ and \mdot\ are likely to be below what is assumed here (and in other works
 on \mdot) because of the saturated luminosity of slim ADs.
 \item
 \Ledd[standard]-based estimates of the continuous growth times of a very large number of BHs 
 are either too long (at the high \Mdot/\mbh\ end)
 or too short (at the low \Mdot/\mbh\ end). Correcting these estimates is essential for improving the understanding
 of BH and galaxy evolution.
 \item
 Some AGN, especially those with the largest mass and lowest \Mdot, must emit significantly below what is assumed
 when using luminosities that are based on bolometric correction factors (\Lbol[standard]). This is true
 even if $\eta$ is the maximum allowed by the thin AD theory.  
 As shown above, this leads to the conclusion that many SDSS sources with such properties contain fast
 spinning BHs.

\end{enumerate}

A recent work by \cite{Wang2013} demonstrates the potential use of slim AGN discs as ``cosmological candles''.
This is based on the idea that above a certain value of \mdot\ ($\sim 1$), the bolometric luminosities of such objects are proportional to the BH mass with only
a very weak dependence on \mdot. These ideas ought to be established theoretically and experimentally. 
Our study, which is based on virial-type estimates of BH masses, indicates that there is
no shortage of such systems at all redshifts up to at least $z=2$. The recent paper by Pu et al. (2013) takes this idea one
step further and shows, by direct reverberation mapping based measurements of \mbh, that such systems exist at low 
redshifts.
It remains to be seen whether this can be extrapolated to higher redshifts and be used to derive distances to active BHs
that are powered by slim ADs.

\section{Acknowledgements}
\label{sec:Acknowledgements}

We thank Jian-Min Wang,  Shai Kaspi and Shane Davis for useful discussions..
This work is supported by an ISF-NSFC grant 83/13.
BT acknowledges the generous support provided by the Benoziyo Center for Astrophysics in the Weizmann Institute of Science and by the Institute for Astronomy in the ETH Z\"{u}rich.


\end{document}